\documentclass[noshowpacs,amsmath,twocolumn,superscriptaddress,8pt,aps,prb]{revtex4-1} % this article uses the Revtex 4-1 format
\usepackage{setspace}
\usepackage{amsmath}
\usepackage{braket}
\usepackage{breqn}
\usepackage{graphicx}
\usepackage[nearskip,margin = 0pt]{subfig}
\usepackage{verbatim} % for commenting
\usepackage{amsfonts}
\usepackage{amssymb}
\usepackage{epstopdf}
\usepackage{physics}
\usepackage{xcolor}
\usepackage{ragged2e}
\usepackage{array}
\usepackage[resetlabels,labeled]{multibib}
\DeclareGraphicsExtensions{.pdf,.eps,.png,.jpg,.mps}

\begin{document}

\title{Optical superradiance of a pair of color centers in an integrated silicon-carbide-on-insulator microresonator}
\author{Daniil M. Lukin$^{*1}$, Melissa A. Guidry$^{*1}$, Joshua Yang$^1$, Misagh Ghezellou$^2$, Sattwik Deb Mishra$^{1}$, Hiroshi Abe$^3$, Takeshi Ohshima$^3$, Jawad Ul-Hassan$^2$, and Jelena Vu\v{c}kovi\'{c}$^{\dagger,1}$\\
\vspace{+0.05 in}
$^1$E. L. Ginzton Laboratory, Stanford University, Stanford, CA 94305, USA.
\\
$^2$Department of Physics, Chemistry and Biology, Link\"oping University, SE-58183, Link\"oping, Sweden
\\
$^3$National Institutes for Quantum Science and Technology, Takasaki, Gunma 370- 1292, Japan
}

\begin{abstract}

An outstanding challenge for color center-based quantum information processing technologies is the integration of 
optically-coherent emitters into scalable thin-film photonics. Here, we report on the integration of near-transform-limited silicon vacancy (V\textsubscript{Si}) defects into microdisk resonators fabricated in a CMOS-compatible 4H-Silicon Carbide-on-Insulator platform. We demonstrate a single-emitter cooperativity of up to 0.8 as well as optical superradiance from a pair of color centers coupled to the same cavity mode. 
We investigate the effect of multimode interference on the photon scattering dynamics from this multi-emitter cavity quantum electrodynamics system.
These results are crucial for the development of quantum networks in silicon carbide and bridge the classical-quantum photonics gap by uniting optically-coherent spin defects with wafer-scalable, state-of-the-art photonics.

\end{abstract}

\maketitle

Color centers\cite{atature2018material, awschalom2018quantum, wolfowicz2021quantum} are among the leading contenders for the realization of distributed quantum information processing, including communication\cite{pompili2021realization, bhaskar2020experimental} and computation\cite{nickerson2014freely}, combining a long-lived multi-qubit spin register\cite{bradley2019ten} with a photonic interface in the solid state. To continue scaling up quantum networks while maintaining high entanglement generation rates, the intrinsically weak interaction between photons and color centers must be enhanced via integration into photonic resonators\cite{sipahigil2016integrated, evans2018photon, lukin20204h, bhaskar2020experimental, crook2020purcell, rugar2021quantum, kuruma2021coupling}. Efforts in cavity integration have already enabled milestone demonstrations such as cavity-mediated coherent interaction between two emitters \cite{evans2018photon}, single-emitter cooperativity exceeding 100 and spin-memory-assisted quantum communication\cite{bhaskar2020experimental}. The ultimate goal of quantum computation and error-protected communication\cite{muralidharan2016optimal} requires the realization of photonic circuits with high complexity and minimal inter-node loss, and will require bringing together all integrated photonics expertise developed in the past two decades.\cite{pelucchi2021potential}

Yet color center technologies cannot at present take advantage of the state of the art in integrated photonics, due to two central challenges. First, thin-film-on-insulator photonics technologies have been incompatible with high-quality color centers: this motivated the focus on bulk-crystal-carving methods\cite{khanaliloo2015high, sipahigil2016integrated, dory2019inverse, wan2020large, babin2021fabrication}, suitable for fabrication of individual devices but restrictive in terms of large-scale monolithic photonic circuits. Second, inversion symmetry, which protects optical transitions from electric fields (to first order\cite{aghaeimeibodi2021electrical, de2021investigation}), had been widely considered to be a prerequisite for color centers to maintain optical coherence in nanophotonic structures. This notion motivates the dominant focus on group-IV color centers in diamond (SiV, SnV, GeV)\cite{bradac2019quantum}, and eliminates from consideration an entire class of materials that lack crystal inversion symmetry. Among these materials is silicon carbide (SiC)\cite{lukin2020integrated}, which has otherwise emerged as the top contender for wafer-scale integration of color centers with excellent spin-optical properties (such as the silicon vacancy (V\textsubscript{Si})\cite{soykal2016silicon, banks2019resonant, morioka2020spin, nagy2021narrow, babin2021fabrication} and the divacancy\cite{anderson2019electrical, bourassa2020entanglement}). 
This inversion symmetry requirement has only recently been challenged in a demonstration of optically-coherent V\textsubscript{Si} in bulk-carved SiC nanobeams\cite{babin2021fabrication}. 

In this work, we demonstrate the integration of optically-coherent non-inversion-symmetric color centers into scalable thin-film SiC nanophotonics. We demonstrate cavity cooperativity of a single V\textsubscript{Si} color center  of up to 0.8, allowing for the observation of dipole-induced transparency\cite{waks2006dipole} in SiC. We achieve a photon detection rate
of up to 0.4~MHz from a single defect into the zero-phonon-line (ZPL), limited by the population shelving in the metastable state. We use this platform to demonstrate superradiant emission of two SiC color centers, and highlight the unique applications of the two-emitter whispering-gallery-mode (WGM) resonator system for quantum information processing architectures. {Our work challenges the notion that inversion symmetry is a prerequisite for nanophotonic integration of optically-coherent spin defects}, and bridges the classical-quantum photonics gap by uniting color centers with CMOS compatible, wafer-scalable, state-of-the-art photonics\cite{song2019ultrahigh, guidry2020optical, guidry2022quantum}. 

\begin{figure*}[t!]
\centering
\includegraphics[width=\linewidth]{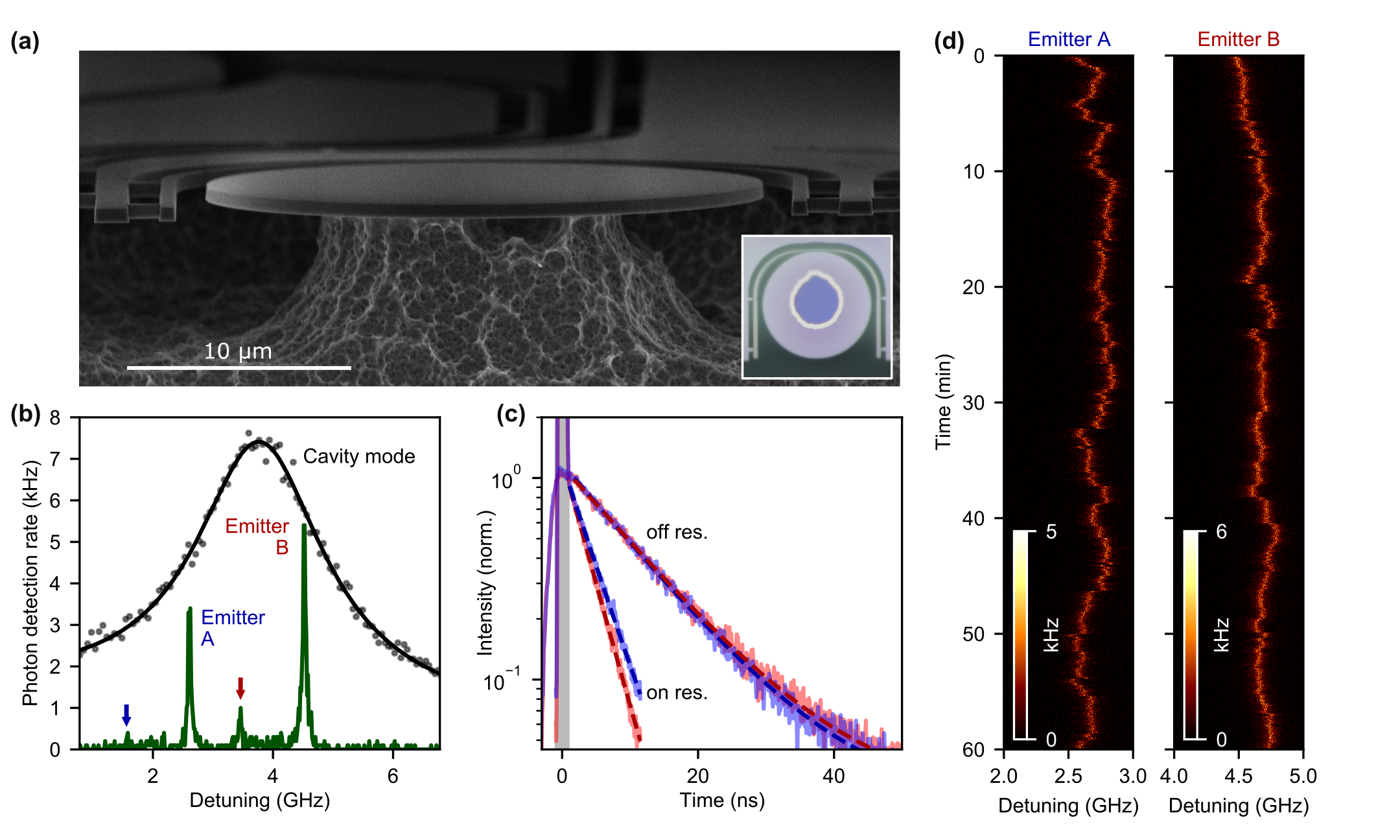}
\captionsetup{format=plain,justification=RaggedRight}
\caption{\textbf{Spectrally-stable V\textsubscript{Si} emitters in integrated 4H-SiCOI photonics.} \textbf{(a)} Scanning electron micrograph of the device. A waveguide, which wraps around the disk (seen in the optical microscope image, inset), is coupled to the resonator. A microscope objective is used to couple light to and from the flat facets of the waveguide. \textbf{(b)} A cavity photoluminescence spectrum (emitter PLE spectrum) in black (green), taken with a scanning resonant laser with 1.5~\textmu W (0.5~pW) of power in the waveguide. We extract a loaded cavity quality factor of Q~$=1.3\cdot10^5$. The prominent peaks at 2.7 and 4.5~GHz detuning are the A\textsubscript{2} transitions of the two emitters. The corresponding A\textsubscript{1} transitions are labelled with arrows. \d{In this figure and henceforth,} laser detuning is relative to 327.113~THz ($916.5$~nm). \textbf{(c)} Lifetime measurements for emitter A (blue) and emitter B (red) on- and off-resonance with the cavity. The gray region represents the excitation pulse. \textbf{(d)} A 1-hour PLE scan of each emitter (while the other is selectively ionized into the dark state), with the cavity positioned on-resonance with the emitter.}
\label{fig:purcell}
\end{figure*}

The photonic device consists of a microdisk resonator integrated with a waveguide (Fig.~\ref{fig:purcell}(a)), fabricated in 4H-Silicon Carbide-on-Insulator (4H-SiCOI)\cite{lukin20204h}. The high-Q transverse-magnetic (TM) modes of the resonator optimally align with the dipole moment of the V\textsubscript{Si} in a \textit{c}-cut wafer\cite{soykal2016silicon}. The coupling waveguide terminates in a flat facet on both ends to allow for efficient single-mode free-space coupling. Details of the fabrication process are presented in the Supplementary Information. We observe a total coupling efficiency from the waveguide to the single-mode fiber of up to 24\%, which includes all setup losses. 
The experiments are performed at 4.3~K in a closed-cycle cryostat (Montana Instruments). The microresonator modes are tuned spectrally via argon gas condensation. A pulsed femtosecond laser centered at 740~nm is used to uniformly excite the emitters in the disk: it couples to all resonator modes simultaneously, owing to its broad spectrum. As the microresonator is gas-tuned, an enhancement of emission at the  V\textsubscript{Si} ZPL wavelength of 916.5~nm (as observed via a spectrometer) indicates Purcell enhancement of one or more V\textsubscript{Si}. With a resonator mode parked at the Purcell enhancement condition, we measure the absorption lines of the coupled emitters via photoluminescence excitation (PLE), where a weak (0.5~pW in the waveguide) continuous-wave resonant laser is scanned across the ZPL while detecting the phonon sideband (PSB) of the emitters. A PLE scan shows that in this device, two emitters are coupled to the cavity (Fig.~\ref{fig:purcell}b), henceforth labeled emitters A and B. The V\textsubscript{Si} is known to feature two spin-preserving optical transitions, A\textsubscript{1} and A\textsubscript{2}, split by 1~GHz \cite{banks2019resonant}. We perform experiments with a weak off-axis external magnetic field that mixes the ground-state spins and eliminates resonant-laser-induced spin-polarization. We focus our study on the A\textsubscript{2} transition of each emitter, which is brighter due to its higher quantum efficiency\cite{banks2019resonant}. We optimize the magnetic field orientation to reduce the relative intensity of the A\textsubscript{1} transition upon resonant driving through coherent population trapping of the spin-$\frac32$ sublevels. Through the absence of a cavity transmission dip, we conclude that the resonator mode is strongly undercoupled to the waveguide. To observe the cavity lineshape, we measure the cavity  photoluminescence noise by scanning across the resonance with higher laser power (1.5~\textmu W in the waveguide) and extract a loaded quality factor of $1.3\cdot10^5$ (Fig.~\ref{fig:purcell}b). 

\begin{figure*}[t!]
\centering
\includegraphics[width=\linewidth]{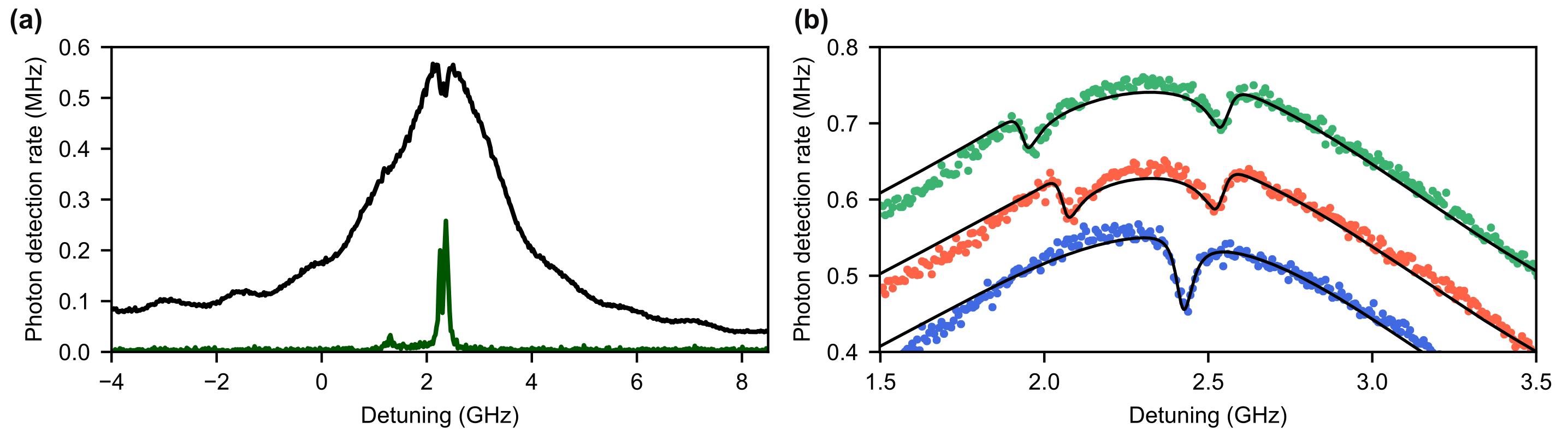}
\captionsetup{format=plain,justification=RaggedRight}
\caption{\textbf{Dipole induced transparency (DIT) in SiC.} \textbf{(a)} A wide laser scan across the cavity resonance, showing the transmission spectrum through the device (black). The V\textsubscript{Si} phonon sideband emission is simultaneously detected (green, multiplied by 50x). Excitation of the resonator mode is performed through a scattering imperfection on the disk edge and transmission through the waveguide is detected. \textbf{(b)} Close-up scan at the cavity center for different emitter detunings. Orange and green traces are offset by +0.1 and +0.2~MHz, respectively.}
\label{fig:DIT}
\end{figure*}

The emitter-cavity coupling rate is a key metric for cavity quantum electrodynamics systems. We determine coupling strength of each emitter to the cavity by measuring the emitter lifetime reduction on resonance, known as Purcell enhancement. First, we selectively ionize one emitter into the dark state via strong resonant excitation, and tune the cavity on-resonance with the remaining bright emitter. We then excite the emitter with 150~ps resonant pulses (obtained via pulse-shaping a mode-locked Ti:Sapphire laser) through the cavity mode and detect the transient ZPL emission using temporal filtering. As shown in Fig.~\ref{fig:purcell}(c), the on-resonance lifetime for emitter A (B) is measured to be 4.2~ns (3.5~ns), which corresponds to a lifetime reduction of 2.7 (3.2) from the bulk lifetime of 11.3~ns \cite{di2021InPreparation}, and a Purcell enhancement $F$ of 28 (37) (see Supplementary Information). From the simulated mode volume of $128(\frac{\lambda}{n})^3$ for the fundamental TM\textsubscript{00} mode, we find the theoretical maximum Purcell enhancement of 77 in this device. The observed Purcell enhancement is comparable to that achieved in the first integrations of the diamond silicon vacancy\cite{sipahigil2016integrated, zhang2018strongly} and tin vacancy\cite{rugar2021quantum, kuruma2021coupling} into photonic crystal nanobeam cavities, despite the much stronger mode confinement of those devices. We attribute this to the optimal dipole overlap of the V\textsubscript{Si} with the cavity TM mode and the less stringent emitter positioning requirements of the microdisks. 
Via resonant pulsed excitation with 1~ns long pulses (generated from a continuous-wave laser using electro-optic amplitude modulation) and detection of the PSB emission with the cavity detuned by $-80$~GHz, we measure the off-resonant lifetime of emitter A (B) to be 10.7~ns (11.1~ns). The minor discrepancy between the off-resonant lifetimes and the bulk lifetime (11.3~ns) is attributed to the coupling of the emitters to other modes of the microdisk.

Although Purcell enhancement has been observed in several color center platforms\cite{crook2020purcell, bracher2017selective, sipahigil2016integrated, zhang2018strongly, rugar2021quantum}, including thin-film diamond\cite{faraon2011resonant} and SiC\cite{lukin20204h}, to date cavity-coupled color centers that retain their optical coherence have only been demonstrated in bulk-carved diamond\cite{sipahigil2016integrated, evans2018photon}. To quantify the optical coherence and the spectral stability of the V\textsubscript{Si} in 4H-SiCOI microdisks, we perform continuous PLE scans on each emitter while on- and off-resonance with the cavity. The on-resonance PLE scans are shown in Fig.~\ref{fig:purcell}(d). Emitters A and B were measured at different times, and the cavity has been centered on the measured emitter before the start of each one-hour acquisition. Over the course of one hour, no emitter ionization is observed, and spectral wandering is below 500~MHz. The average single-scan optical transition linewidth for emitter A (B) is found to be 54.3(3)~MHz (63.4(3)~MHz), which corresponds to 17~MHz (18~MHz) of spectral diffusion beyond the transform limit. Repeating the measurement off-resonance, we find the emitter A (B) linewidth to be 37.8(8)~MHz (38.5(8)~MHz), which corresponds to 24~MHz of spectral diffusion beyond the transform limit (see Supplementary Information). The reduced spectral diffusion on-resonance may be due to a decreased rate of excitation of surface-related defects, because the well-confined TM cavity mode is efficiently excited with low laser power. These results indicate excellent spectral stability of the nanophotonics-integrated V\textsubscript{Si}.

\begin{figure*}[t!]
\centering
\includegraphics[width=\linewidth]{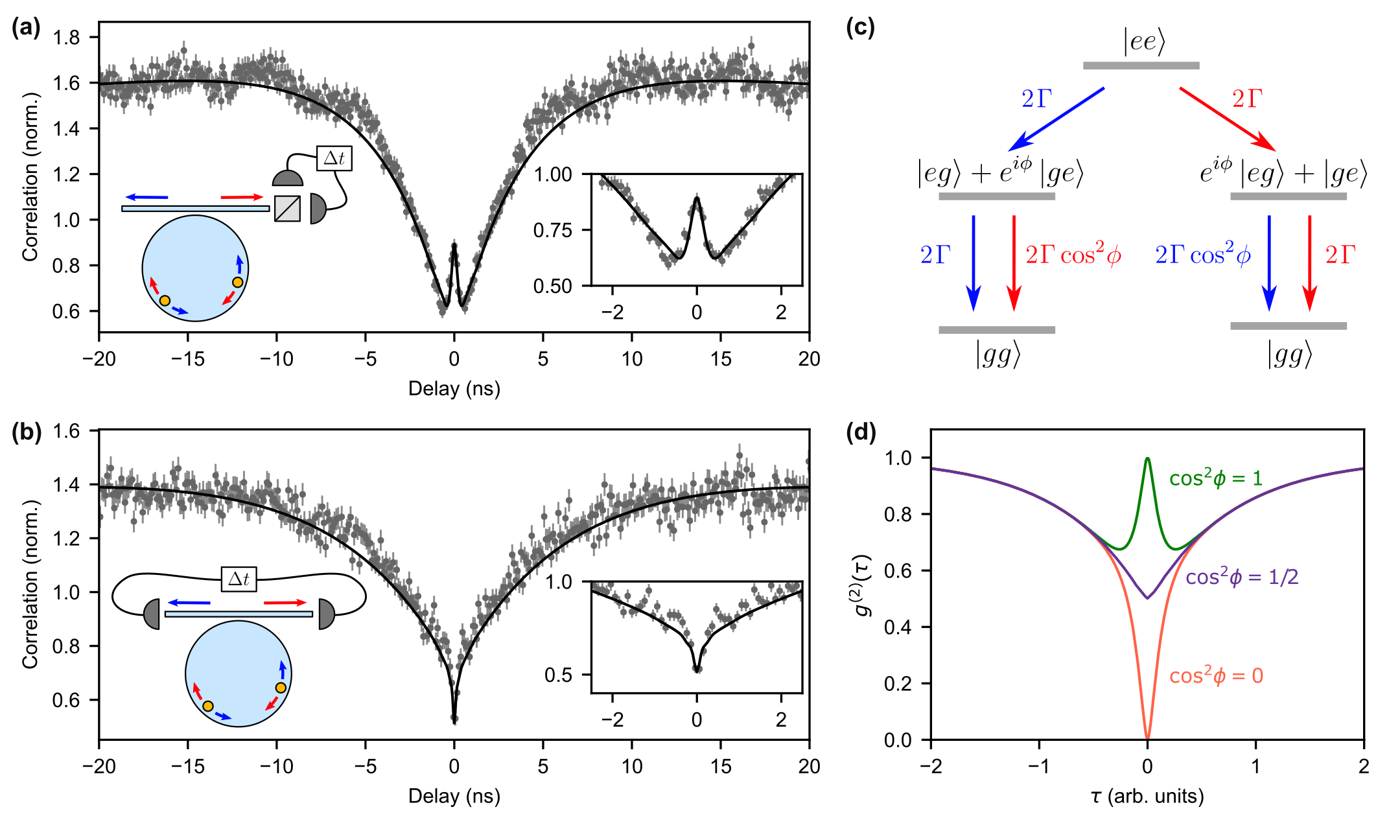}
\captionsetup{format=plain,justification=RaggedRight}
\caption{\textbf{Superradiant emission of two V\textsubscript{Si} color centers.} \textbf{(a)} Second-order correlation of the photon emission along one waveguide direction displays bunching at zero time delay, a signature of superradiance. Inset: zoom-in of the superradiance feature. \textbf{(b)} The relative phase $\phi$ of the emitters impacts the cross-correlation photon statistics between the opposite waveguide directions and can produce anti-bunched emission. The solid line in a,b is the numerical fit based on a five-level model\cite{di2021InPreparation} of the V\textsubscript{Si}.  \textbf{(c)} The level structure representing the pair of two-level-system emitters decaying into degenerate clockwise (red arrows) and counterclockwise (blue arrows) optical modes. The corresponding transition rates are indicated next to the arrows, where $\Gamma$ is the unmodified single-emitter decay rate into a propagating mode. \textbf{(d)} Theoretically-predicted phase-dependent cross-correlation between clockwise and counterclockwise modes for a pair of ideal two-level emitters. }
\label{fig:g2}
\end{figure*}
From the measured Purcell enhancement and off-resonant emitter linewidths, we calculate the emitter-cavity cooperativity $C = \frac{4 g^2}{\kappa \gamma }$ to be 0.6 and 0.8 for emitter A and B, respectively (see Supplementary Information). This regime enables the observation of dipoled-induced transparency (DIT)\cite{waks2006dipole}, where the V\textsubscript{Si} scatter photons from an input coherent state. Because the device studied here is strongly under-coupled to the bus waveguide, DIT is difficult to observe through waveguide transmission. We instead excite the disk through a scattering point on its edge, and detect emission into the waveguide, thus in effect performing the measurement in a drop-port configuration\cite{waks2006dipole} (see Supplementary Information). Scanning the continuous-wave laser across the disk resonance, DIT dips for both emitters are clearly observed, shown in Fig.~2(a,b). The slow spectral drift of the emitters allows us to measure DIT for different relative detunings. Looking forward, spin initialization, targeted emitter placement, and cavities with a larger $Q/V$ metric \cite{lukin20204h, song2019ultrahigh} will enable stronger transmission contrast in DIT for the realization of spin-photon entanglement and spin-readout via the modification of cavity reflectivity \cite{evans2018photon, bhaskar2020experimental}.

Photon interference between two color centers, a prerequisite for the generation of remote spin-spin entanglement, has been an outstanding challenge in silicon carbide. Here, we demonstrate two-photon interference between two microdisk-integrated emitters, which arises from their collective coupling to the same cavity mode. To observe photon interference in the continuous wave regime, an above-resonant laser is coupled to a resonator mode around 730~nm to excite both emitters. We note that while above-resonant excitation in bulk crystal  has been used to obtain nearly transform-limited photon emission from the V\textsubscript{Si}\cite{morioka2020spin}, we observe that in nanostructures it induces rapid spectral diffusion due to disturbance of the surface charge environment, broadening the optical linewidths to approximately 0.5~GHz. This spectral instability reduces the rate of superradiant emission (however, optical coherence may be preserved using resonant excitation, as shown later in the work). Fig.~3(a) shows the second-order auto-correlation $g^{(2)}(\tau)$ of the color centers' collective emission in the Hanbury Brown and Twiss configuration, where emission into the waveguide is split between two detectors via a beamsplitter. The sharp peak at zero time delay is a signature of superradiant emission and the probabilistic generation of entanglement between the two color centers. This feature has also been observed with up to three waveguide-integrated quantum dots \cite{kim2018super, grim2019scalable} and a pair of waveguide-integrated silicon vacancy centers in diamond\cite{sipahigil2016integrated, machielse2019quantum}. In contrast, for cross-correlations between the two waveguide propagation directions, an anti-bunching interference dip is observed (Fig.~3(b)). This feature is indicative of photon pairs preferentially leaving the resonator in the same direction.

\begin{figure*}[t!]
\centering
\includegraphics[width=0.85\linewidth]{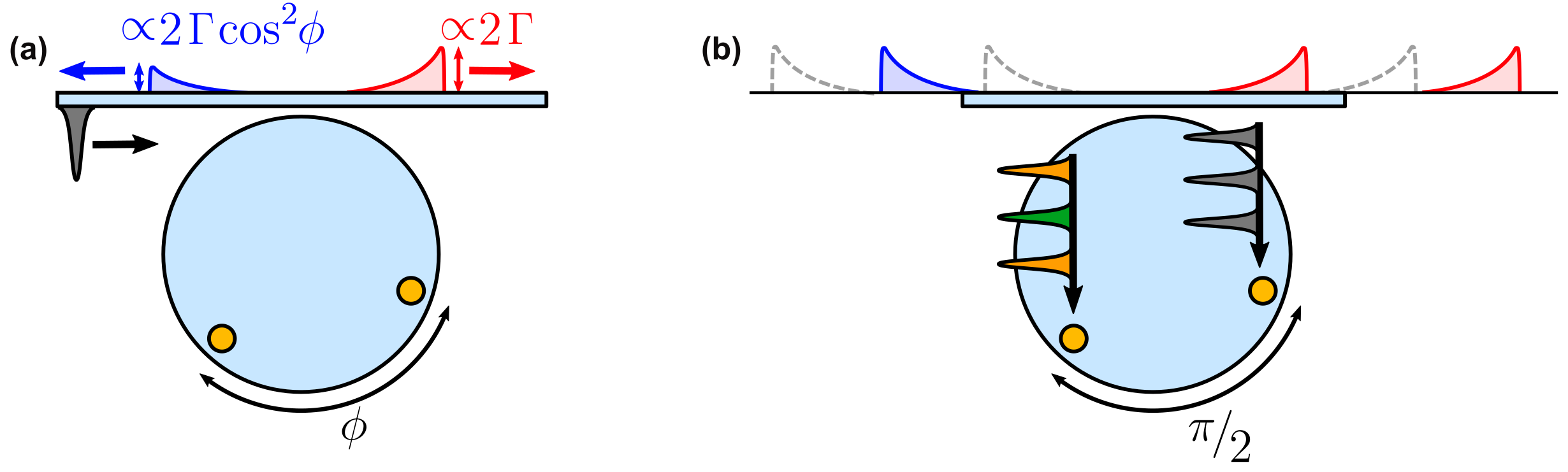}
\captionsetup{format=plain,justification=RaggedRight}
\caption{\textbf{Manipulating the single-photon emission of a pair of emitters.} \textbf{(a)} Weakly exciting the emitters with a resonant pulse (grey) through the CW mode will prepare the system in the superposition $(e^{i\phi}\ket{eg} + \ket{ge})/\sqrt{2}$, which will result in asymmetric emission rates. \textbf{(b)} By independently controlling the excitation phase of the two emitters positioned such that $\phi=\pi/2$, the microresonator incorporates the functionality of a single-photon router. The phase of the free-space excitation pulse is represented by the color, where green, grey,  and orange correspond to $\pi/2$, $0$, and $-\pi/2$, respectively.}
\end{figure*}

The experimentally-observed photon statistics are explained by the out-of-phase coupling of the two emitters to a pair of degenerate clockwise and counterclockwise optical modes of the resonator. The interaction Hamiltonian for this system can be written as
\begin{equation}
    H_I = g_A \sigma_A^\dag S_A + g_B \sigma_B^\dag S_B + \text{h.c.},
    \label{eq:H_I}
\end{equation}

where $\sigma_A$ and $\sigma_B$ are the lowering operators for emitters A and B, respectively, and $g_A$ and $g_B$ are the emitter-cavity coupling strengths; each emitter couples to its own standing wave supermode $S_A = (a_{\text{CW}} + a_{\text{CCW}})/\sqrt{2}$ and $S_B = (e^{-i\phi} a_{\text{CW}} + e^{i\phi} a_{\text{CCW}})/\sqrt{2}$, where $a_{\text{CW}}$ ($a_{\text{CCW}}$) is the clockwise (counterclockwise) resonator propagating mode, and phase $\phi$ corresponds to the emitters' azimuthal separation in the resonator. 
Consider two special cases: (i) for $\phi=(0\mod\pi)$, $S_A = \pm S_B$ and the two emitters couple to the same standing wave mode, resulting in a single-mode interaction\cite{evans2018photon}; (ii) for $\phi=(\pi/2 \mod \pi)$, $S_A$ and $S_B$ are orthogonal, and in the standing wave basis the emitters are de-coupled. However, because the measurement is performed in the propagating mode basis $\{a_{\text{CW}},a_{\text{CCW}} \}$ (corresponding to emission to the right and to the left, respectively), the pair of emitters exhibits interference for all values of $\phi$. For $(\phi\mod\pi) \neq 0$, the cross-correlation between the two waveguide propagation directions will reveal interference features unique to a multi-mode, multi-emitter system. 

The collective emission behavior can be understood via a cascaded decay diagram shown in Fig.~3(c). Starting with the two-emitter excited state $\ket{ee}$, emission into the clockwise mode projects the emitters into the superposition state $(e^{i\phi}\ket{eg} + \ket{ge})/\sqrt{2}$. From this state, decay via clockwise emission proceeds with the superradiant rate $2\Gamma$, where $\Gamma$ is the unmodified single-emitter decay rate into a propagating mode. In contrast, the rate of counterclockwise emission is modified by $\cos^2\phi$, as follows from the transition amplitude ${\bra{gg}(e^{i\phi}\sigma_A + \sigma_B)(e^{i\phi}\ket{eg}+\ket{ge})/\sqrt{2}}$. When $\cos^2\phi=0$, photons leave the resonator always in the same direction, which corresponds to perfect antibunching in the cross-correlation. For $\cos^2\phi=\pm1$, the cross-correlation is identical to the autocorrelation on a single waveguide direction. These cases are illustrated in Fig.~3(d). The correlation measurements (Fig.~3(a,b)) are fit to a reduced five-level emitter model\cite{di2021InPreparation} with free parameters of excitation power, $\phi$, cavity detuning, and background noise. The presence of background noise from the above-resonant excitation reduces the interference contrast.

\begin{figure*}[t!]
\centering
\includegraphics[width=0.8\linewidth]{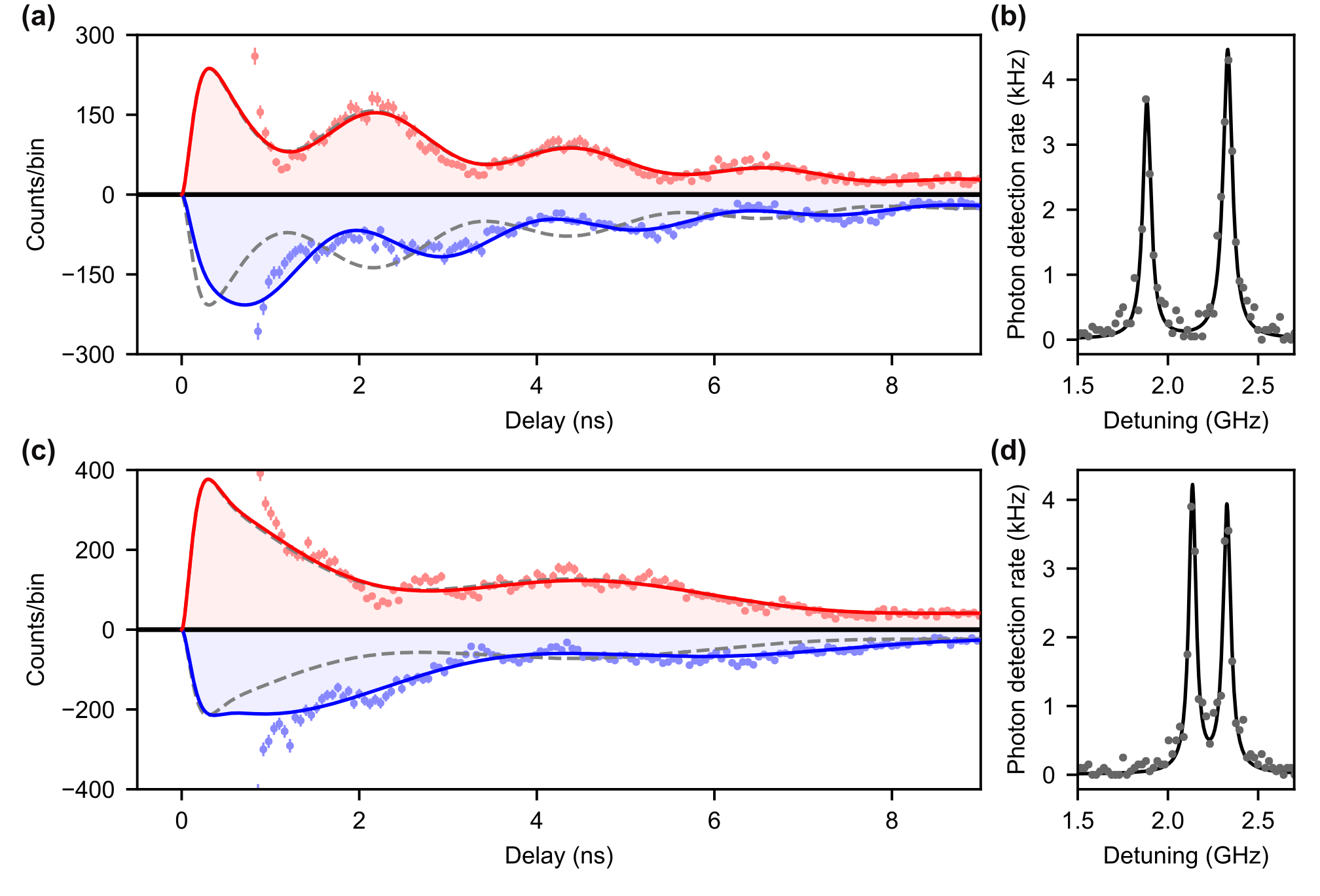}
\captionsetup{format=plain,justification=RaggedRight}
\caption{\textbf{Chiral single-photon scattering from a pair of emitters in a WGM resonator} \textbf{(a)} The emitter pair is excited through the CW mode. Photons scattered into the CW (red) and CCW (blue) mode are time-correlated to the excitation pulse, tracing out the temporal shape of the emitted single-photon wavepacket. The solid red and blue lines represent the simulated expectation values $\langle a^\dagger_{\text{CW}}a_{\text{CW}} \rangle $  and $\langle a^\dagger_{\text{CCW}}a_{\text{CCW}}\rangle$, respectively. The asymmetric CW and CCW emission arises from non-trivial emitter phase difference, inferred to be $(0.34\pi \mod \pi)$. The simulated case where ${\phi=0}$ is shown as a grey dotted curve, in which case the emission is symmetric. \textbf{(b)} The PLE spectrum of the two emitters shows frequency separation of 0.44~GHz, which is used as a fixed parameter in the simulation of the wavepacket in (a). \textbf{(c,d)} Same as panels (a,b) but for emitter frequency separation of 0.19~GHz, with inferred phase $\phi = (0.28\pi \mod \pi)$.}
\label{fig:single_photon_interference}
\end{figure*}
\textbf{}

Excitation of emitters via above-resonant optical fields increases spectral diffusion and is compatible neither with spin-selective excitation nor optical coherent control. To overcome this, we use resonant excitation to coherently manipulate the two-emitter superposition in the single photon subspace, a regime which has enabled pioneering quantum network experiments with NV centers in diamond \cite{humphreys2018deterministic, pompili2021realization}. Consider exciting the two emitters through the waveguide via fast resonant pulses in the clockwise direction (Fig.~4(a)). In the bad-cavity regime ($\kappa \gg \gamma$) and with resonator finesse $\mathcal{F} \gg 1$, if the two emitters are initially in the ground state $\ket{gg}$, a pulse instantaneously prepares the system into a superposition state
\begin{multline}
\ket{\psi} = (1 - P_e)\ket{gg} \\ + \sqrt{P_e(1 - P_e)}(e^{i\phi} \ket{eg} + \ket{ge} ) + P_e e^{i\phi} \ket{ee},
\end{multline}
where $P_e$ is the single-emitter excitation probability. In the weak excitation limit ($P_e \ll 1$), the probability of double-excitation is negligible, and the system is prepared in the superposition  $(e^{i\phi}\ket{eg} + \ket{ge})/\sqrt{2}$ conditioned on the detection of a scattered photon. This corresponds to the preparation of the two-emitter system into the intermediate level of the diagram in Fig.~3(c). The emission from this state will proceed superradiantly in the clockwise direction independent of $\phi$, but the back-scattering rate will be modified by $\cos^2\phi$ (Fig.~4(a)).  For $\phi = \pi/2$, complete directionality is achieved. This is analogous to classical chiral scattering observed in WGM resonators coupled to a pair of dielectric\cite{peng2016chiral} and plasmonic\cite{cognee2019cooperative} nanostructures. The $\phi = \pi/2$ condition can be used to implement routing of single photons from an emitter pair (Fig.~4(b)) and, as shown in the Supplementary Information, enables efficient spin-spin entanglement protocols. We note that due to the cavity-mediated coupling of emitters, collective scattering of input light is strengthened in the high-cooperativity regime, whereas in a waveguide system high-cooperativity emitters will act as individual strong scatterers.

The combination of preserved optical coherence and spectral stability enables the experimental realization of single-photon interference between two V\textsubscript{Si} emitters, shown in Fig.~5. Because the emitters' transitions are not degenerate, their relative phase will precess at the rate equal to their frequency difference, which is observed as an oscillation in the single-photon wavepacket. Notably, the phase difference in the oscillations of CW and CCW emission originates from the relative emitter phase $\phi$. As described in the Supplementary Information, the non-unity contrast of the oscillations is due to the distribution of the spin population across the bright spin-$\frac{1}{2}$ and dark spin-$\frac{3}{2}$ manifolds (corresponding to the optical transitions A\textsubscript{2} and A\textsubscript{1}, respectively). The oscillations persist throughout the entire wavepacket, confirming nearly transform-limited photon emission, essential for interference-based entanglement generation\cite{humphreys2018deterministic}. The smaller amplitude oscillation with a 1~ns period is due to incomplete suppression of the A\textsubscript{1} emission line of the V\textsubscript{Si} (Fig.~1(b)). The free parameters in the numerical model are cavity detuning, $\phi$, and the population of the spin-$\frac{1}{2}$ manifold. From the numerical fit, the spin population is inferred to be unpolarized, as expected in an off-axis magnetic field. The relative emitter phase inferred from the data in Fig.~5(a) and 5(c) is $\phi = (0.34\pi \mod \pi)$ and $\phi = (0.28\pi \mod \pi)$, respectively. We note that $\phi$ was observed to drift in time, attributed to nonuniform deposition of water ice on the resonator as a result of the asymmetric resonator undercut geometry. This explanation is consistent with the observed slow systematic drift of emitter spectral separation by approximately 1~GHz per day due to strain from the ice. 

The present study of the cavity quantum electrodynamics of a color center pair in a microdisk resonator suggests that despite a typically lower quality factor-to-mode volume ratio (Q/V) than high-confinement photonic crystal cavities, WGM resonators may offer unique capabilities and warrant further consideration for applications in chip-integrated quantum information processing. As we show in the Supplementary Information, the two-emitter chiral scattering at the $\phi=\pi/2$ condition enables efficient entanglement generation via single-photon interference. %Alternatively, microdisk integration of a color center with angular-momentum-non-conserving optical transitions may enable
{Through integration with a color center that exhibits two orthogonal circularly polarized transitions such as the divacancy in silicon carbide, the microdisk could be used to realize single-emitter chiral light-matter interaction\cite{scheucher2016quantum,lodahl2017chiral}.}
Furthermore, the microdisk is a promising platform for near-term many-body quantum optics demonstrations with solid state spins, as many individually-addressable and spectrally-tunable emitters may be integrated into a single resonator\cite{lukin2020integrated}. 

In conclusion, we have demonstrated near-unity cooperativity between a color center and a microresonator fabricated in a wafer-scalable, CMOS-compatible semiconductor photonics platform. Additionally, we observe two-photon superradiance and single-photon interference between two SiC color centers. 
The integration of V\textsubscript{Si} into state-of-the-art microring resonators\cite{guidry2022quantum} and high-confinement photonic crystal cavities\cite{song2019ultrahigh} would enable deterministic emitter-photon interactions in SiC.
Taken together with the recent demonstrations of nuclear spin control\cite{bourassa2020entanglement,babin2021fabrication}, wide spectral tuning via electric fields\cite{anderson2019electrical, lukin2020spectrally} and single-shot readout\cite{anderson2021five}, silicon carbide satisfies the prerequisites to implement a fully-monolithic quantum photonic processor. The maintained spin-optical coherence of the V\textsubscript{Si} at elevated temperatures of up to 20~K \cite{babin2021fabrication, udvarhelyi2020vibronic} offers an additional degree of flexibility for operation with low-cost cryogenic systems. 
Finally, the spectral stability of the V\textsubscript{Si}, despite its substantial dipole moment\cite{lukin2020spectrally}, suggests that a first-order insensitivity to electric fields is not a prerequisite for color center compatibility with nanostructures; this result motivates a continued effort toward the integration of other SiC color centers, such as the divacancy\cite{anderson2019electrical}, the nitrogen-vacancy center\cite{von2016nv}, the vanadium center\cite{spindlberger2019optical,wolfowicz2020vanadium}, and the chromium ion\cite{diler2020coherent}, into nanophotonics.

$^*$These authors contributed equally \\

$^\dagger$ jela{\makeatletter @\makeatother}stanford.edu
\\

\noindent\textbf{Acknowledgments}
\noindent  
We gratefully acknowledge discussions with Florian Kaiser, Charles Babin, Di Liu, J{\"o}rg Wrachtrup, \"{O}ney O. Soykal, Daniel Riedel, Christopher P. Anderson, Ki Youl Yang, Shahriar Aghaeimeibodi, Alexander D. White. This work is funded in part by the U.S. Department of Energy (DoE), Office of Science, under Awards DE-SC0019174, DE-Ac02-76SF00515, and the National Quantum Information Science Research Centers. D.M.L. acknowledges the J. Hewes Crispin and Marjorie Holmes Crispin Stanford Graduate Fellowship (SGF) and the National Defense Science and Engineering Graduate (NDSEG) Fellowship. M.A.G. acknowledges the Albion Hewlett SGF and the NSF Graduate Research Fellowship. J.Y. acknowledges the NDSEG. J.U.H. acknowledges support from Swedish Research Council (grant No. 2020-05444), Knut and Alice Wallenberg Foundation (grant No. KAW 2018-0071), and the EU H2020 project QuanTELCO (grant No. 862721). T.O. acknowledges grants  JSPS KAKENHI 20H00355 and 21H04553. Part of this work was performed at the Stanford Nanofabrication Facility (SNF) and the Stanford Nano Shared Facilities (SNSF).

\bibliography{Reference}
\clearpage

%\appendix 
\renewcommand{\thesection}{\Roman{section}}
\setcounter{section}{0}

\end{document}

\title{Supplementary Information: Optical superradiance of a pair of color centers in an integrated silicon-carbide-on-insulator microresonator}
\author{Daniil M. Lukin$^{*1}$, Melissa A. Guidry$^{*1}$, Joshua Yang$^1$, Misagh Ghezellou$^2$, Sattwik Deb Mishra$^{1}$, Hiroshi Abe$^3$, Takeshi Ohshima$^3$, Jawad Ul-Hassan$^2$, and Jelena Vu\v{c}kovi\'{c}$^{\dagger,1}$\\
\vspace{+0.05 in}
$^1$E. L. Ginzton Laboratory, Stanford University, Stanford, CA 94305, USA.
\\
$^2$Department of Physics, Chemistry and Biology, Link\"oping University, SE-58183, Link\"oping, Sweden
\\
$^3$National Institutes for Quantum Science and Technology, Takasaki, Gunma 370- 1292, Japan
}

%\appendix 
\renewcommand{\thefigure}{S\arabic{figure}}
\renewcommand{\thesection}{\Roman{section}}
\renewcommand{\bibnumfmt}[1]{[S#1]}
\renewcommand{\citenumfont}[1]{S#1}
\setcounter{figure}{0}
\setcounter{section}{0}

\maketitle

\onecolumngrid 

\tableofcontents

\newpage

\section{Device fabrication}

\begin{figure*}[h]
\centering
\includegraphics[width=\textwidth]{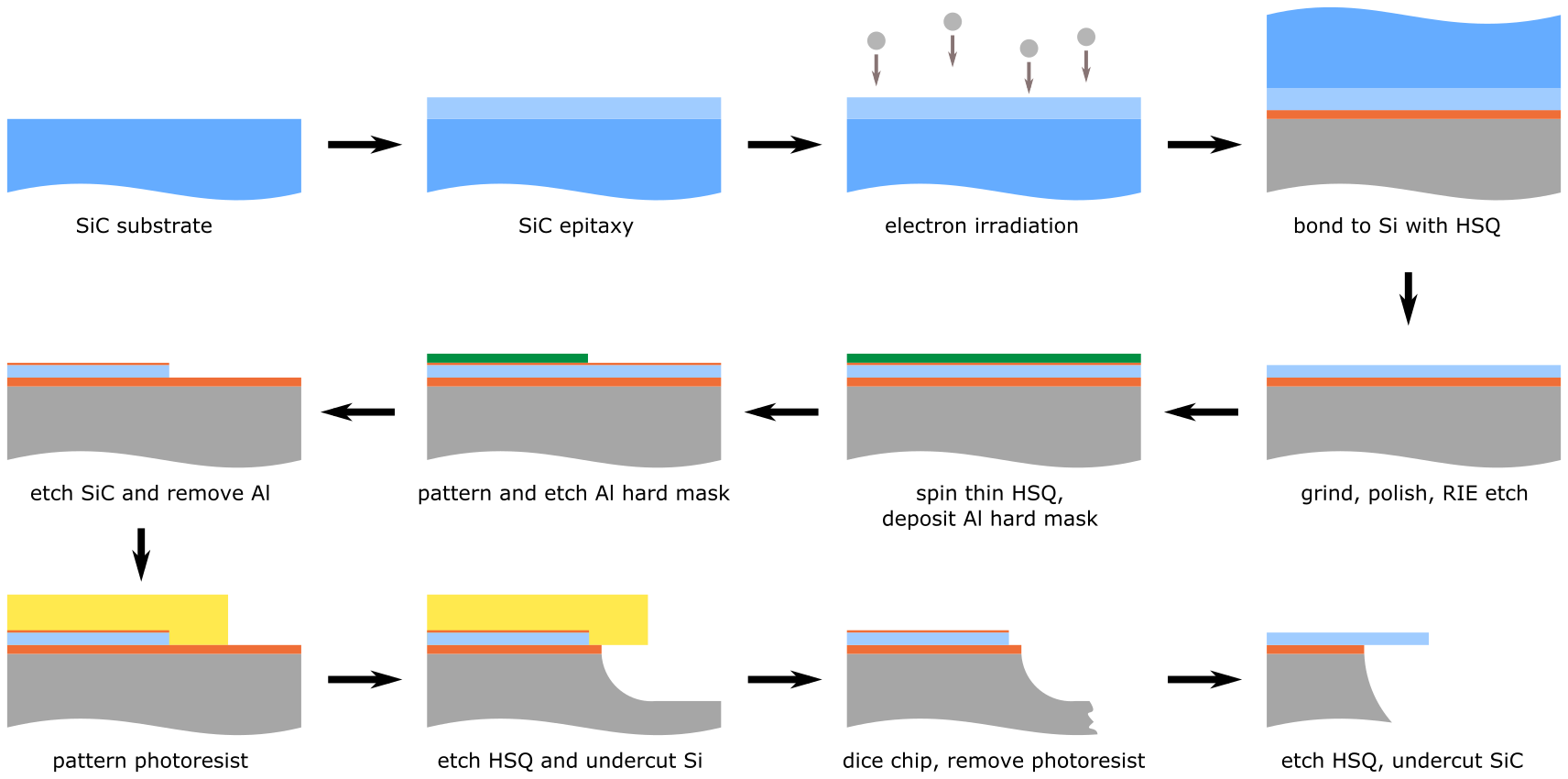}
\captionsetup{format=plain,justification=RaggedRight}
\caption{\textbf{Device fabrication process flow.} Colors correspond to materials as follows. Blue: SiC substrate. Light blue: SiC epitaxy. Grey: Si substrate. Orange: HSQ. Green: hardmask. Yellow: photoresist.}
\label{fig:fabrication}
\end{figure*}

The device fabrication process is summarized in Fig.~\ref{fig:fabrication}. A 20~\textmu m n-doped (nitrogen concentration $2\cdot 10^{13}$~cm\textsuperscript{-3}) SiC epilayer is grown by chemical vapour deposition on a n-type (0001) 4H-SiC substrate. The SiC is irradiated with 2~MeV electrons with a fluence of $1\cdot10^{13}$~cm\textsuperscript{-2} to generate V\textsubscript{Si} defects. The SiC is bonded to a Si substrate via an HSQ flowable oxide layer (FOx-16, Dow Corning) and annealed at 550~$^\circ$C for 2~hours to strengthen the bond and activate V\textsubscript{Si} defects. The SiC is then thinned via grinding, polishing, and reactive-ion etching (RIE) \cite{lukin20204h} to 450-620~nm. A 50~nm protective layer of HSQ is spun, followed by e-beam evaporation of the etch hard mask layer (5~nm Ti / 155~nm Al / 5~nm Ti). The device geometry is patterned via e-beam lithography (JEOL 6300-FS) in ZEP520A resist (Zeon Corp), and transferred into the Al hardmask layer via chlorine-based RIE. The SiC layer is etched using SF\textsubscript{6} in a capacitively-coupled plasma etcher (Oxford Plasmalab 100) at an etch rate of 45~nm/min, with gas flow rate of 50~sccm, pressure of 7~mTorr, etch power of 100~W, and substrate temperature maintained at 20~$^\circ$C. For best quality of the waveguide facet used for in- and out-coupling, it is defined as part of the lithography and RIE etching together with the rest of the geometry, to avoid rough facets that can result from dicing a waveguide.  After the SiC etch, the Al hardmask is then removed via a wet etch in Aluminum Etchant Type A. The final steps of the fabrication achieve an undercut device diced in close proximity to the waveguide facet. This is done as follows: First, an approximately 50~\textmu m wide and 10~\textmu m deep Si trench is created 15~\textmu m from the waveguide ends using photoresist mask and XeF\textsubscript{2} isotropic Si etch. Then, the chip is diced along the trench, while the photoresist provides protection to the device layer. Finally, the photoresist is removed, and the device layer is uniformly undercut via wet HF etch and XeF\textsubscript{2} gas etch, to suspend the resonator and waveguide.

\newpage
\section{Experimental setup}
\begin{figure*}[h]
\centering
\includegraphics[width=\textwidth]{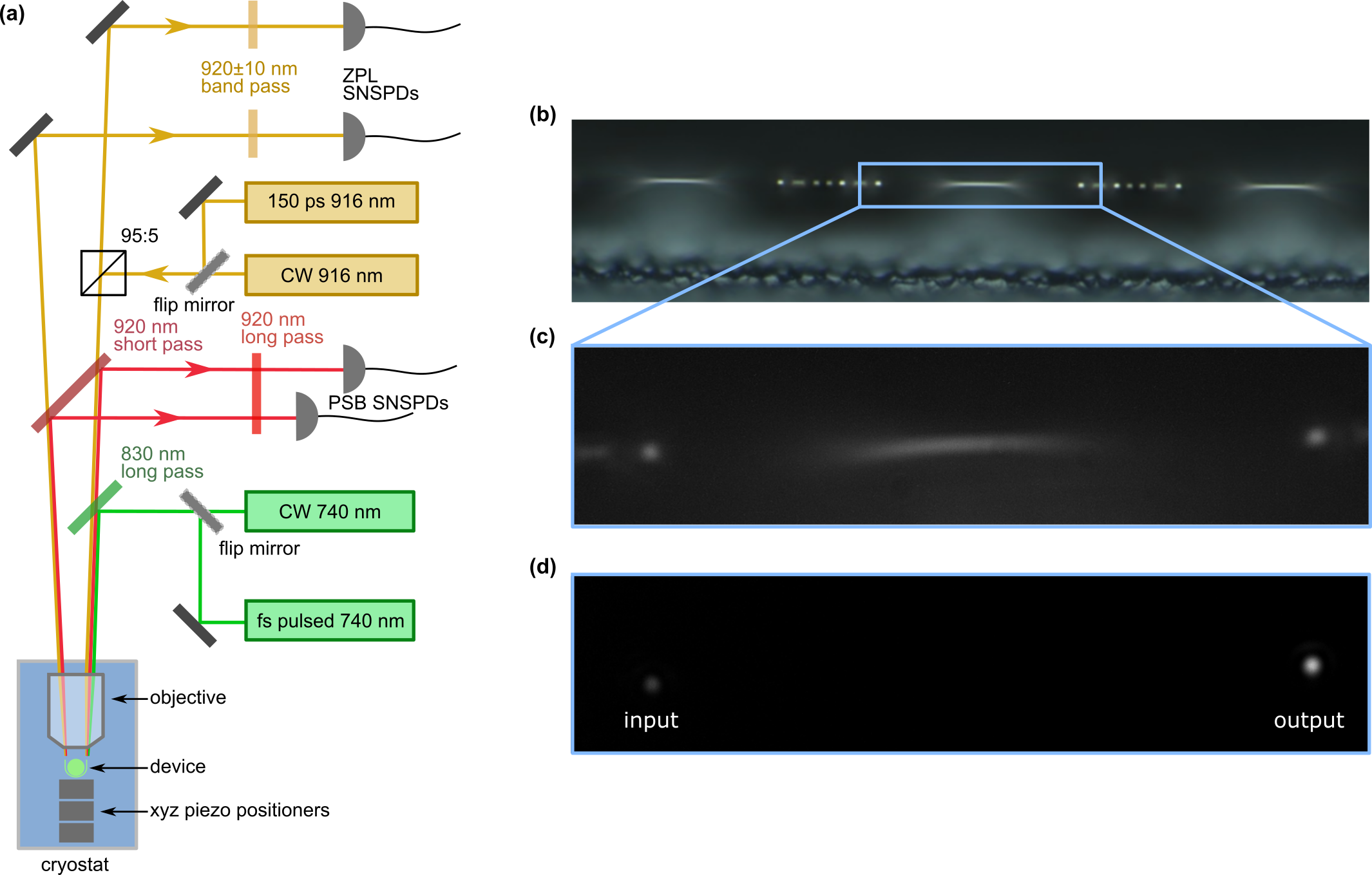}
\captionsetup{format=plain,justification=RaggedRight}
\caption{\textbf{Experimental setup.} \textbf{(a)} Diagram of optical paths, laser sources, and detectors.  \textbf{(b)} Optical microscope image of a row of disk resonators (three resonators are visible), taken using a commercial optical microscope. \textbf{(c)} Optical image of a single device under illumination as seen through the cryostat objective. \textbf{(d)} Optical image of the device without illumination and laser light coupling into the left waveguide facet, passing through the waveguide and emitting from the right waveguide facet.}
\label{fig:setup}
\end{figure*}

The experimental setup is shown schematically in Fig.~\ref{fig:setup}(a). The sample is mounted in a closed-cycle cryostat with the cryo-optic module (Montana Instruments) where an objective with an NA of 0.9 is mounted inside the vacuum chamber. The sample is mounted on a three-axis piezo positioner stack (Attocube) so that the waveguide facets point up toward the objective (optical images shown in Fig.~\ref{fig:setup}(b-d)). The optical paths coupling to the two waveguide ends are spatially separated into separate fiber couplers. Dichroic mirrors allow for simultaneous collection of ZPL and PSB emission \cite{bernien2012two}. For continuous-wave above-resonant excitation (such as to measure two-photon interference) and for V\textsubscript{Si} charge control, a continuous-wave Ti:Sapphire laser is used, with wavelength tuned to couple to a resonator mode around 740~nm for uniform excitation of the resonator mode volume. During the gas-tuning phase, a femtosecond pulsed Ti:Sapphire laser is used to achieve multi-mode excitation of the microresonator that does not vary with resonance shifting due to gas deposition. For resonant excitation, a continuous-wave Ti:Sapphire laser is used for PLE and DIT measurements, whereas a picosecond pulsed Ti:Sapphire laser is used for on-resonance lifetime reduction and single-photon interference measurements. The picosecond laser outputs 5-15~ps FWHM pulses, which are sent to a pulse-shaper to produce 150~ps pulses that are bandwidth matched to the microresonator optical mode that the V\textsubscript{Si} are coupled to. Photons are detected using superconducting nanowire single photon detectors (SNSPDs), produced by PhotonSpot Inc., and photon correlations are processed with the TimeTagger Ultra (Swabian Instruments). A limitation of the current experimental configuration, where the sample is mounted on the side to collect photon emission from the waveguide end, is the lack of optical access to the top of the resonator, precluding individual excitation of the emitters with free-space beams. Access from above would also make possible spatially-resolved laser-ablation of the condensed gas used to tune the cavity, which would enable fine control over the relative emitter phase $\phi$.

\newpage

\section{Emitter linewidths on- and off-resonance}
\begin{figure}[h]
\centering
\includegraphics[width=0.95\textwidth]{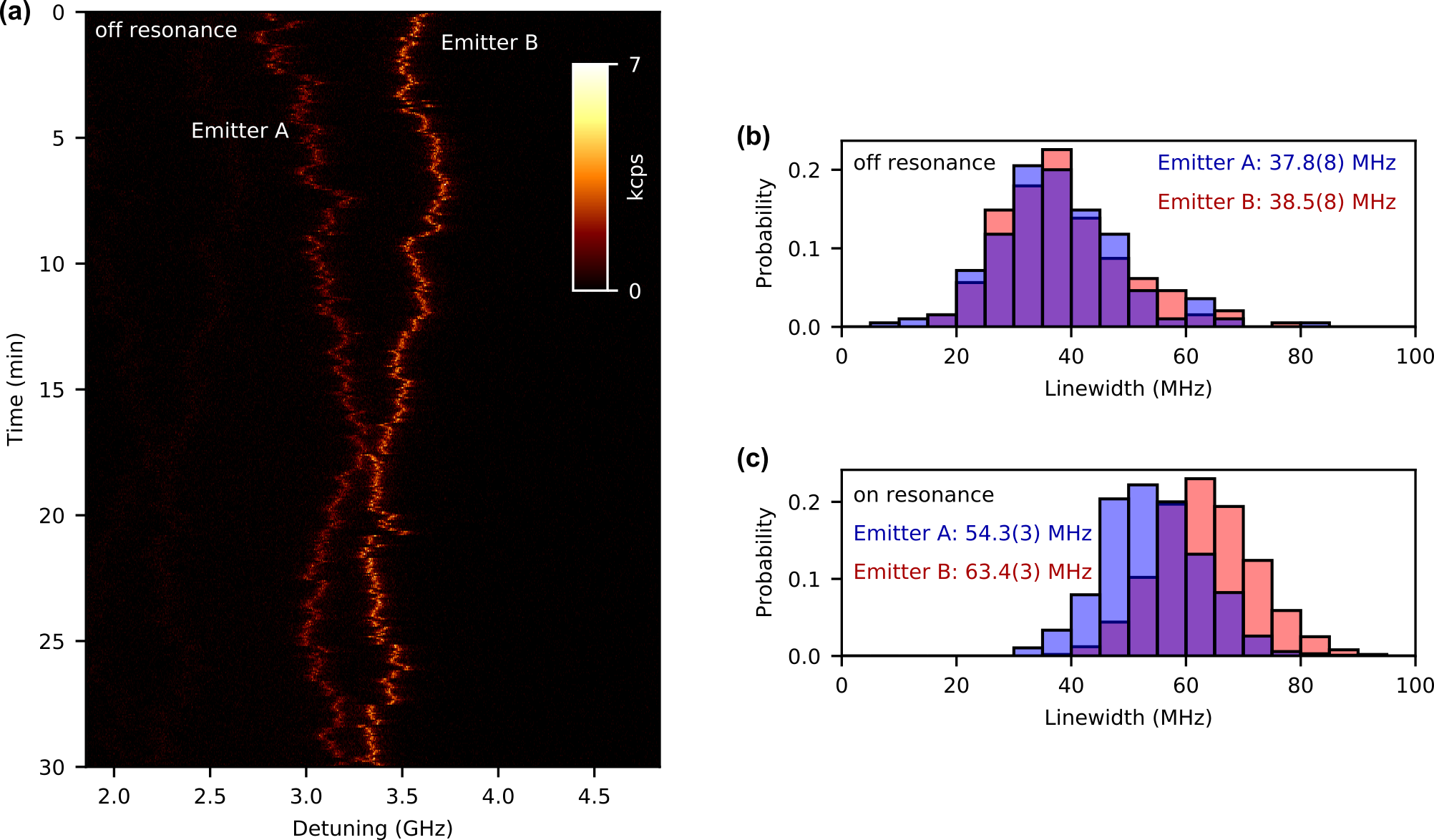}
\captionsetup{format=plain,justification=RaggedRight}
\caption{\textbf{Emitter linewidths on- and off-resonance with the cavity.} (a) A continuous PLE scan of the two emitters with the cavity far-detuned. (b) A histogram of fitted single-scan linewidths. Indicated in the figure is the mean fitted linewidth and its standard error. (c) Histogram of time-averaged scans for PLE data presented in Fig.~1(d) of the main text, showing spectral broadening caused by lifetime reduction of the optical transition. Emitter B transition is broader due to the stronger Purcell enhancement.}
\label{fig:linewidths}
\end{figure}

In order to determine the emitter-cavity cooperativity, the rates of pure dephasing and emitter-cavity coupling on resonance must be known. For the most reliable cooperativity estimate, we measure the emitters' linewidths both on-resonance with the cavity, as well as when the cavity is far-detuned. The series of on-resonance PLE scans is shown in Fig.~1 of the main text. The series of off-resonance PLE scans over the course of 30 minutes is shown in Fig.~\ref{fig:linewidths}(a). The distributions of the fitted linewidths for the off- and on-resonance PLE scans are shown in Fig.~\ref{fig:linewidths}(b) and (c), respectively. The pure dephasing rate on- and off-resonance, as inferred from the difference between the measured mean linewidth and the transform limit, is similar in both cases, 17 MHz (18 MHz) for emitter A (B) on-resonance and 24~MHz off-resonance. For the estimate of cooperativity, the larger dephasing rate (off-resonant) is used.

\newpage

\section{Gas tuning and saturation of photon detection rate}

\begin{figure}[h]
\centering
\includegraphics[width=0.45\textwidth]{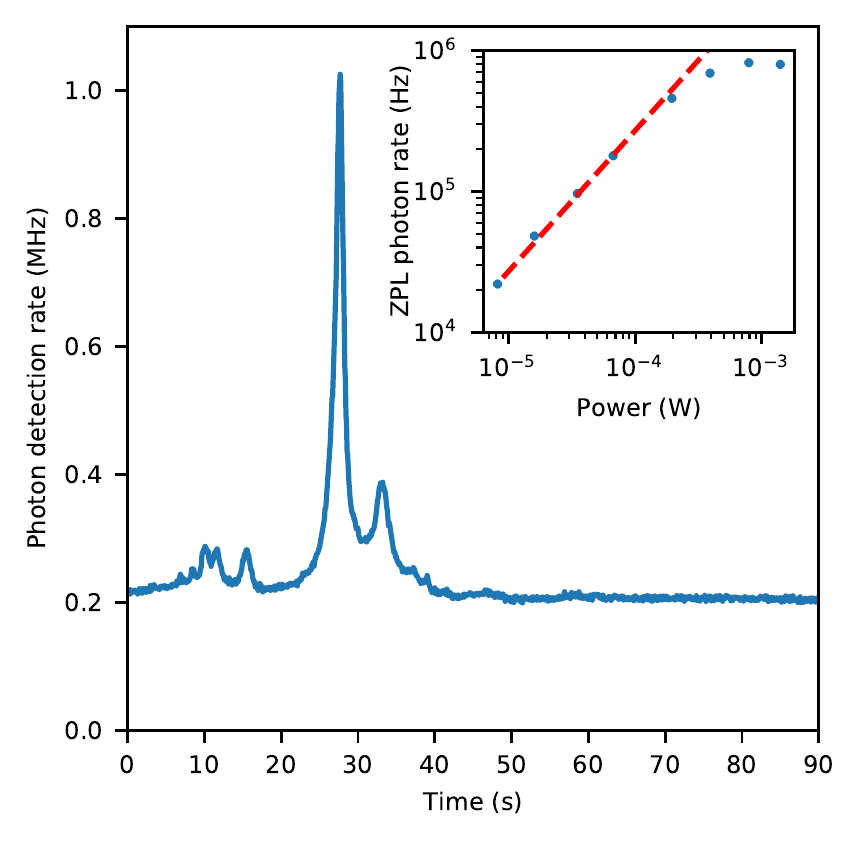}
\captionsetup{format=plain,justification=RaggedRight}
\caption{\textbf{Gas tuning and saturation of photon detection rate.} Photon detection rate in a 20~nm spectral window (910-930~nm) around the V\textsubscript{Si} ZPL during continuous red-tuning of the microdisk resonance wavelength via gas condensation. The Purcell enhancement condition is observed as a sharp peak in time. Smaller peaks correspond to weaker coupling to other detuned emitters. Excitation is performed with a 730~nm, 80~MHz repetition rate femtosecond laser (0.79~mW power measured before the objective). Inset shows background-subtracted peak ZPL photon detection rate for varying laser power.}
\label{fig:purcell_vs_time}
\end{figure}

A representative gas-tuning spectrum upon above-resonant excitation with an 80~MHz fs laser, without narrowband spectral filtering of emission is shown in Fig.~\ref{fig:purcell_vs_time}. The high background photon rate arises because the entire volume of the disk resonator has to be excited in order to excite the two emitters, due to the lack of free-space optical access to the disk. Selective excitation of emitters from above via free-space optical beams would drastically reduce background fluorescence. The background-subtracted ZPL detection rate from the two emitters at saturation is 0.8~MHz, corresponding to a single-emitter ZPL detection rate of 0.4~MHz in the case of equal coupling, and higher in reality due to unequal cavity-coupling rates (denoted $g_A$ and $g_B$ in main text).

\vspace{-1em}
\section{Purcell enhancement and cooperativity calculation}
The Purcell factor of the emitter-cavity system is defined as the ratio of the rate of emission into the cavity to the unmodified ZPL decay rate. Using the radiative lifetime of the A\textsubscript{2} transition of 15.9~ns, \cite{di2021InPreparation} and DWF of 8.5\%\cite{udvarhelyi2020vibronic, shang2020local} from the literature, we conclude that the unmodified ZPL rate of the A\textsubscript{2} transition is $1/186.5$~ns$^{-1}$. From the non-radiative rate from the A\textsubscript{2} transition of 40~ns$^{-1}$, we obtain the on-resonance cavity emission rate for emitter A (B) of $1/5.1$~ns$^{-1}$ ($1/6.7$~ns$^{-1}$). From this, we obtain Purcell enhancement of 28 and 37 for emitter A and B, respectively.

The cooperativity of the emitter-cavity system is given by
\begin{equation}
    C = \frac{4 g^2}{\kappa \gamma } \equiv \frac{\Gamma}{\gamma}
    \label{eq:cooperativity}
\end{equation}
where $g$ is the single-photon Rabi frequency, $\kappa$ is the cavity decay rate, $\gamma$ is the total decay rate of the emitter, and $\Gamma$ is the rate of emission into the cavity. In Fig.~S3, we measure averaged off-resonant linewidths of $\gamma_A/{2\pi} = 37.8$~MHz and $\gamma_B/{2\pi} = 38.6$~MHz for the two emitters. In Fig.~1(c) of the main text, we measure the on-resonant optical transition lifetimes to be $\tau_A = 4.2$~ns and $\tau_B = 3.5$~ns.  The lifetime of the A\textsubscript{2} optical transition of the V\textsubscript{Si} in bulk crystal is known to be $\tau_0 = 11.3$~ns\cite{di2021InPreparation}. We infer $\Gamma_A/{2\pi} = 23.8$~MHz and $\Gamma_B/{2\pi} = 31.4$~MHz  from the relation $1/\tau_i = \Gamma_i + 1/\tau_0$ and using Eq.~\ref{eq:cooperativity} calculate cooperativities for the two emitters of  $C_A = 0.6$ and $C_B = 0.8$. Using the measured $\kappa/{2\pi} = 2.8$~GHz, we determine  $g_A/{2\pi} = 125$~MHz and $g_B/{2\pi} = 150$~MHz.

\newpage

\section{Spin selective temporally-filtered resonance fluorescence}
\begin{figure}[h]
\centering
\includegraphics[width=\textwidth]{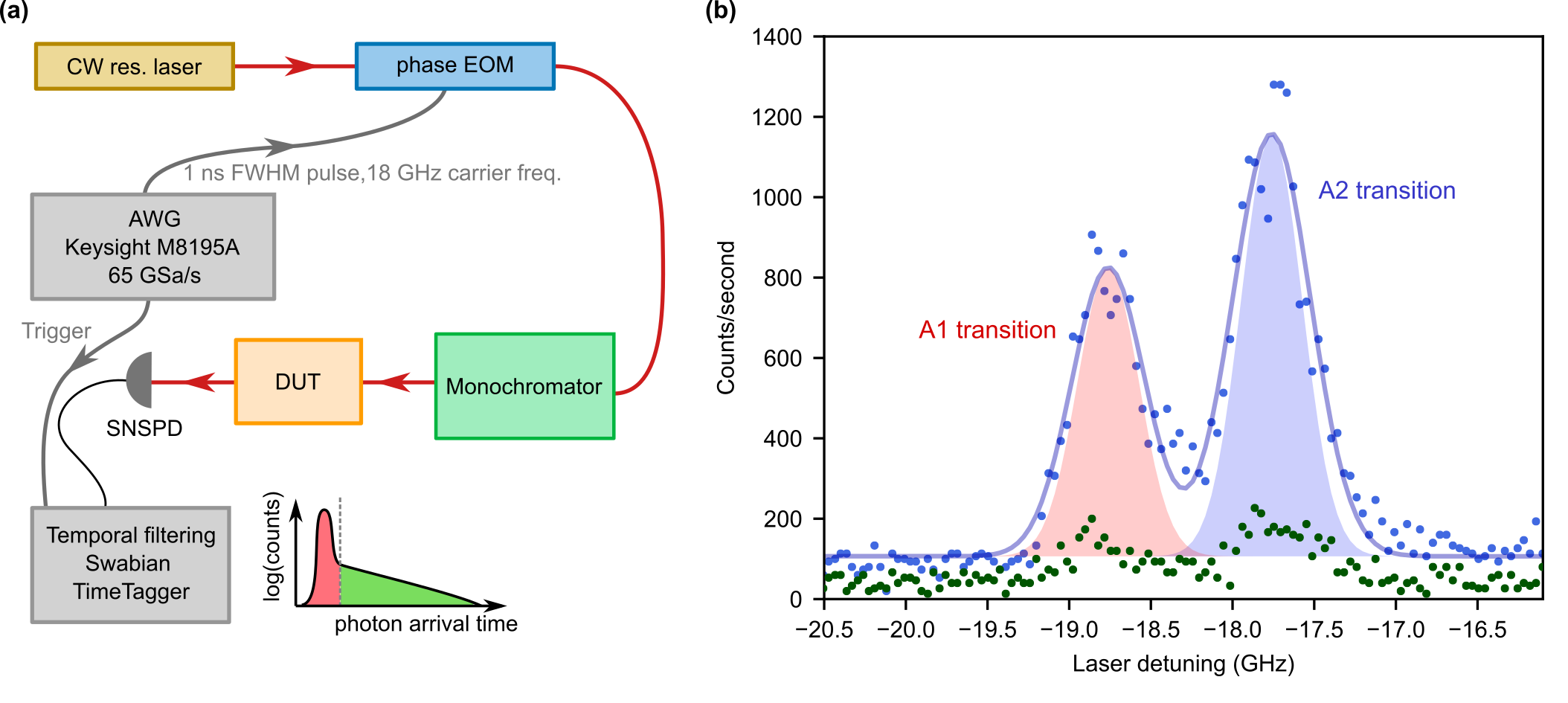}
\captionsetup{format=plain,justification=RaggedRight}
\caption{\textbf{Spin selective temporally-filtered resonance fluorescence.} \textbf{(a)} Diagram of the experimental configuration. Temporally-modulated laser sideband at 18~GHz is generated using a phase electro-optic modulator (EOM) driven by an arbitrary signal generator (AWG). The sideband is spectrally filtered and sent to the device. The detected photons arrival times are correlated with the excitation pulse (Swabian Time Tagger): The earlier photon arrivals corresponding to the excitation pulse are discarded. \textbf{(b)} Resonance fluorescence spectrum of a single V\textsubscript{Si} (blue data points) taken with 1~ns FWHM excitation pulses. The shaded areas correspond to the excitation pulse transform limit (0.44~GHz FWHM). The green data points are the simultaneously-acquired phonon side-band emission. Due to the strong Purcell enhancement of the defect, the phonon side-band detection rate is significantly lower than that of the ZPL.}
\label{fig:spin_selective_pulsed_excitation}
\end{figure}

In the main text, pulsed resonant excitation to detect transient ZPL emission was performed with picosecond pulses from a mode-locked laser expanded to 150~ps via pulse shaping. With this approach, however, it is difficult to implement selective excitation of just one of the transitions, due to their small separation of 1~GHz. An alternative approach to generate spectrally narrower pulses is via electrooptic modulation. We use an electro-optic \textit{phase} modulator combined with spectral filtering to electronically define the optical pulse shape via an arbitrary waveform generator. Generating pulses thus, rather than via an electo-optic \textit{amplitude} modulation, achieves high rejection ratio (60~dB) and is insensitive to environmental fluctuations, not requiring any active stabilization of the modulator. Figure~\ref{fig:spin_selective_pulsed_excitation}(a) details the experimental configuration. Using 1~ns FWHM pulses, corresponding to a 0.44~GHz FWHM in frequency, we perform temporally-filtered resonance-fluorescence on a single V\textsubscript{Si}, observing well resolved A\textsubscript{1} and A\textsubscript{2} transitions, shown in  Figure~\ref{fig:spin_selective_pulsed_excitation}b.

\newpage

\begin{figure}[h]
    \centering
    \includegraphics[width=0.4\textwidth]{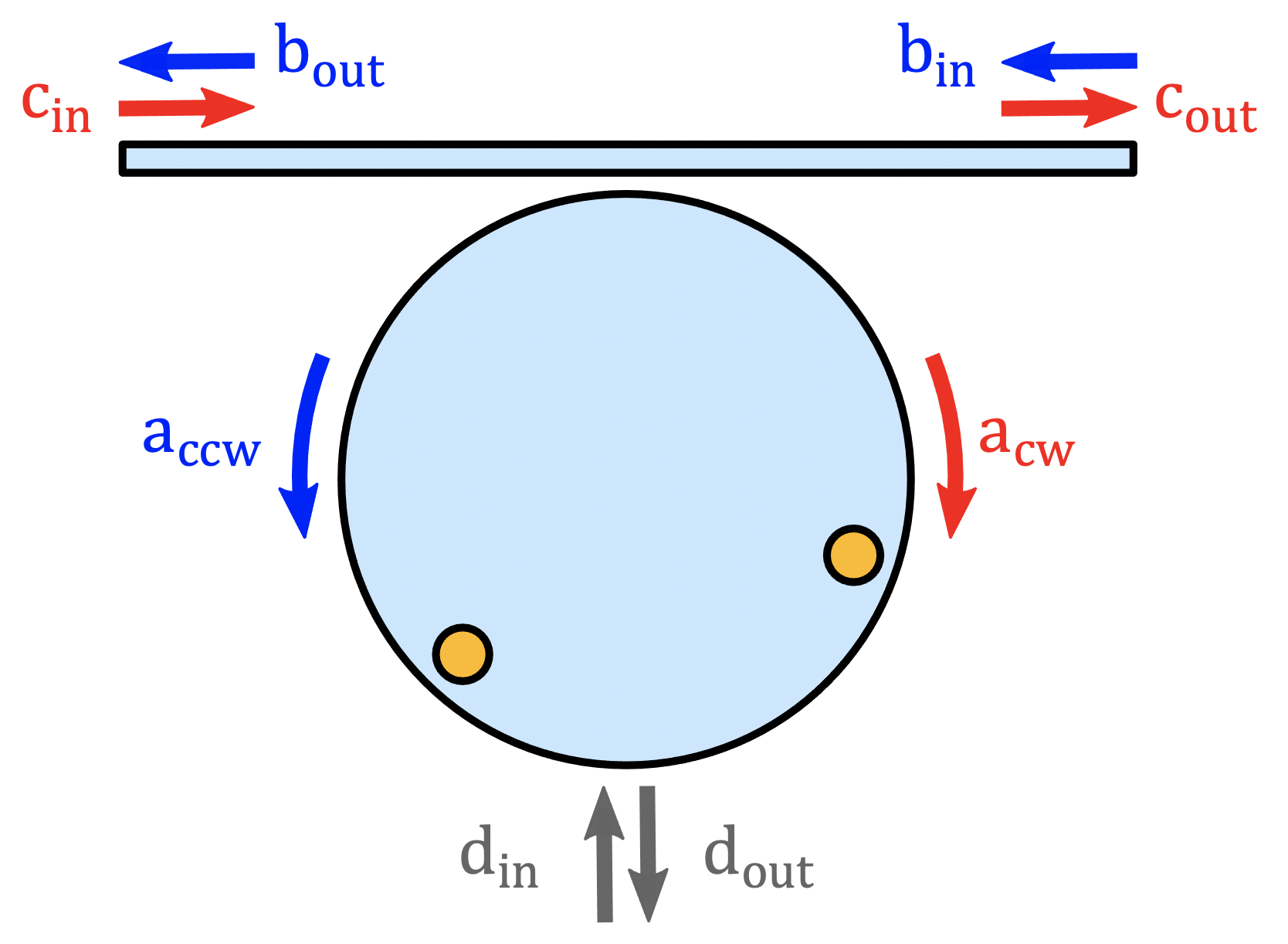}
    \caption{Bath and cavity operators for a disk resonator coupled to a single bus waveguide along with a scattering port.}
    \label{fig:DIT_WGM}
\end{figure}

\section{Dipole induced transparency in a whispering gallery mode resonator}
Dipole induced transparency is often explored for a system where a single cavity mode couples to one or more emitters\cite{waks2006dipole}, which is the case for a nanobeam\cite{sipahigil2016integrated}, two-dimensional photonic crystal\cite{song2019ultrahigh}, or Fabry-P\'{e}rot resonator\cite{riedel2017deterministic}. The whispering gallery mode (WGM) resonator is distinct in that each resonance corresponds to two degenerate cavity modes. In this section, we define a model to describe transmission through the WGM resonator in the presence of quantum emitters. We describe our model in the clockwise (CW) and counter-clockwise (CCW) propagating mode basis. These modes are degenerate with resonance frequency $\omega_0$ and are described by cavity mode annihilation operators $a_{\text{CW}}$ and $a_{\text{CCW}}$. Using the input-output formalism\cite{GardinerInputOutput1985}, we write the relations between the bath operators, including the scattering-defect excitation port described in the main text (see Figure~\ref{fig:DIT_WGM}):
\begin{align*}
    b_{\text{out}} &=   \sqrt{\kappa_c} a_{\text{CCW}} + b_{\text{in}} \\
    c_{\text{out}} &=   \sqrt{\kappa_c} a_{\text{CW}} + c_{\text{in}} \\
    d_{\text{out}} &=   \sqrt{\kappa_d} D + d_{\text{in}} = \sqrt{\kappa_d} ( \sqrt{\alpha} \cdot a_{\text{CW}}  + \sqrt{1-\alpha} \cdot a_{\text{CCW}} ) + d_{\text{in}} 
\end{align*}

Here, $D$ is the resonator mode which the scattering port couples to fully. Note that it is not necessarily a standing wave: $D$ couples with strength $\alpha$ to $a_{\text{CW}}$ and $(1-\alpha)$ to $a_{\text{CCW}}$. We have defined two coupling rates: $\kappa_c$ between the resonator and the bus waveguide and $\kappa_d$ between the resonator and the scattering channel. The emitters in the WGM resonator couple maximally to two different standing waves:
\begin{align*}
    S_1 &= \frac{1}{\sqrt{2}} [a_{\text{CW}} e^{i\theta} + a_{\text{CCW}} e^{-i\theta} ]\\
    S_2 &= \frac{1}{\sqrt{2}} [a_{\text{CW}} e^{i\phi} + a_{\text{CCW}} e^{-i\phi} ]
\end{align*}
where $\theta$ and $\phi$ define the azimuthal orientation (phase) of the standing waves relative to the excited mode $D$. The emitters, modeled as a pair of two-level systems with associated annihilation operators $\sigma_1$ and $\sigma_2$, couple to these standing waves with coupling coefficients $g_1$ and $g_2$.  We write a non-Hermitian Tavis-Cummings Hamiltonian which includes the decay of the emitters and the cavity:
\begin{align*}
H_{\text{TC}} = &(\omega_0 + \Delta - \delta/2 - i \gamma_1 ) \sigma^\dag_1 \sigma_1 + (\omega_0 + \Delta + \delta/2 - i \gamma_2 ) \sigma^\dag_2 \sigma_2 \\
&+ (\omega_0 - i \kappa) (a_{\text{CW}}^\dag a_{\text{CW}} + a_{\text{CCW}}^\dag a_{\text{CCW}} ) + \underbrace{\Big[ g_{1} S_1^\dag \sigma_1 +  g_{2} S_2^\dag \sigma_2 + \text{h.c.} \Big]}_{H_I}
\end{align*}
where $\omega_0$ is the cavity resonance frequency, $\Delta$ is the frequency difference between the cavity and the center of the two emitters, $\delta$ is the frequency difference between the two emitters, $\gamma_j$ is the linewidth of the $j$th emitter (which includes all sources of decay and dephasing), and $\kappa$ is the total decay rate of the cavity. We can explicitly write out the coupling term with respect to the CW and CCW modes:
$$ g_{1} S_1^\dag \sigma_1 +  g_{2} S_2^\dag \sigma_2 = \frac{g_1}{\sqrt{2}}( e^{-i \theta} a_{\text{CW}}^\dag \sigma_1 + e^{i\theta} a_{\text{CCW}}^\dag \sigma_1 ) + \frac{g_2}{\sqrt{2}}( e^{-i \phi} a_{\text{CW}}^\dag \sigma_2 + e^{i\phi} a_{\text{CCW}}^\dag \sigma_2 ) $$
We can define coupling coefficients:
$$G_1 = \frac{g_1}{\sqrt{2}} e^{-i\theta}, \hspace{1em} G_2 = \frac{g_2}{\sqrt{2}} e^{-i\phi}$$
and re-write the interaction term:
$$H_I = ( G_1 a_{\text{CW}}^\dag + G_1^* a_{\text{CCW}}^\dag ) \sigma_1 + ( G_2 a_{\text{CW}}^\dag + G_2^* a_{\text{CCW}}^\dag ) \sigma_2 + \text{h.c.}$$
The Heisenberg equations for the two CW and CCW cavity modes are defined as 
\begin{align*}
\dot{a}_{\text{CW}} &= - i[H_{TC}, a_{\text{CW}}] - \frac{\kappa}{2} a_{\text{CW}} - \sqrt{\kappa_c} c_{\text{in}} - \sqrt{\alpha} \sqrt{\kappa_d} d_{\text{in}}\\
\dot{a}_{\text{CCW}} &= - i[H_{TC}, a_{\text{CCW}}] - \frac{\kappa}{2} a_{\text{CCW}} - \sqrt{\kappa_c} b_{\text{in}} - \sqrt{1-\alpha} \sqrt{\kappa_d} d_{\text{in}}
\end{align*}
Then we can write all four Heisenberg equations (in the frequency domain):
\begin{align*}
-i\omega {a}_{\text{CW}} &= (-i\omega_0 - \frac{\kappa}{2} ) a_{\text{CW}} - \sqrt{\kappa_c} c_{\text{in}} - \sqrt{\alpha} \sqrt{\kappa_d} d_{\text{in}} - i G_1 \sigma_1 - i G_2 \sigma_2 \\
-i\omega {a}_{\text{CCW}} &= (-i\omega_0 - \frac{\kappa}{2} ) a_{\text{CCW}} - \sqrt{\kappa_c} b_{\text{in}} - \sqrt{1-\alpha} \sqrt{\kappa_d} d_{\text{in}} - i G_1^* \sigma_1 - i G_2^* \sigma_2\\
-i\omega \sigma_j &= -i[ (\omega_0 + \Delta + (-1)^j \frac{\delta}{2}) - \frac{\gamma_j}{2} ] \sigma_j - i G_j^* a_{\text{CW}} - i G_j a_{\text{CCW}}
\end{align*}
The measurement of Figure~2 of the main text describes the transmission through the ``drop'' waveguide formed between the input scattering point on the disk and the output bus waveguide: 
\begin{align*}
    t_c &= \langle c_{\text{out}} \rangle / \langle d_{\text{in}} \rangle\\
t_b &= \langle b_{\text{out}} \rangle / \langle d_{\text{in}} \rangle
\end{align*}
We solve for $a_{\text{CW}}$ in terms of the input bath operators and use the expectation values of our input-output equations:
\begin{align*}
    \langle c_{\text{in}} \rangle &=  \langle b_{\text{in}} \rangle = 0\\
    \langle c_{\text{out}} \rangle &= \sqrt{\kappa_c} \langle a_{\text{CW}} \rangle\\
    \langle b_{\text{out}} \rangle &= \sqrt{\kappa_c} \langle a_{\text{CCW}} \rangle
\end{align*}
We solve the system of equations to arrive at the following expressions:
$$t_c = \frac{\Gamma_1 \Gamma_2 \sqrt{\kappa_d \kappa_c} }{\Phi^2 - \psi^+ \psi^- } [\sqrt{1-\alpha} \psi^+ -\sqrt{\alpha} \Phi  ] $$
$$t_b = \frac{\Gamma_1 \Gamma_2 \sqrt{\kappa_d \kappa_c} }{\Phi^2 - \psi^+ \psi^- } [\sqrt{\alpha} \psi^- -\sqrt{1-\alpha} \Phi ] $$
where we have defined
$$\Gamma_1(\omega) = [ -i(\omega - \omega_0 - \Delta + \delta/2 ) + \gamma_1 / 2 ], \hspace{1em} \Gamma_2(\omega) = [ -i(\omega - \omega_0 - \Delta - \delta/2 ) + \gamma_2 / 2 ]$$
$$ \psi^+ =  {G_1^2}{\Gamma_2} +  {G_2^2}{\Gamma_1} , \hspace{1em} \psi^- =  {G_1^{*2}}{\Gamma_2} +  {G_2^{*2}}{\Gamma_1} $$
$$\Phi  = (-i(\omega-\omega_0) + \kappa/2) \Gamma_1 \Gamma_2  + {|G_1|^2 \Gamma_2} +  {|G_2|^2\Gamma_1} $$
To account for the Fano shape observed experimentally in the transmission spectrum, we add a coherent term with a defined phase $\rho$ and amplitude $B$. We include the relative amplitude $A$ with an offset $C$:
$$T_c(\omega) =   | A \cdot t_c(\omega) + B \cdot  e^{i \rho } |^2 + C$$
Note that this equation does not account for the non-unity occupation probability\cite{zhang2018strongly} of the spin-1/2 ground state (corresponding to the A\textsubscript{2} transition). This results in an underestimate of the coupling strength between the excited mode $D$ and the emitter standing waves $S_1$ and $S_2$. In the main text, the fits are performed as follows. The cavity and Fano parameters $\omega_0$, $\kappa$, $B$, $C$, and $\rho$ are fit to the wide scan data (Fig.~2(a)). These parameters are fixed for all other fits. For each close-in scan in Fig.~2(b), the PLE measurement which is taken simultaneously is fit to extract the parameters $\delta$ and $\Delta$. We set the values for $g_1$ and $g_2$ to those extracted from the lifetime measurements. In the DIT fit, the free parameters are $\theta$, $\phi$,  $\alpha$, and $A$.

\section{Two-emitter single photon interference modeling}

The modeling of the single-photon temporal envelope shown in Fig.~5 of the main text is performed using QuTiP based on a simplified three-level model, where the bright ground state spin-$\frac12$ manifold is treated as one state, $\ket\uparrow$, coupled to one excited state, $\ket e$. This is an appropriate approximation because all optical transitions within the spin-$\frac{1}{2}$ manifold are near-degenerate and so the fine structure does not impact the photon emission. Additionally, a third state is introduced $\ket\downarrow$, which represents the dark spin population which is not excited by the optical excitation pulse. This model thus includes the effect of the dark spin population on the interference.

The first step is to calculate the two-emitter state upon the -- assumed instantaneous -- weak coherent excitation. Before the application of a weak optical excitation pulse, the two-emitter system is in a mixed state
$$\rho_0 = (1-P_B)^2\dyad{\downarrow\downarrow} + P_B^2\dyad{\uparrow\uparrow} + P_B(1-P_B)(\dyad{\uparrow\downarrow} + \dyad{\downarrow\uparrow}),$$
where $P_B$ is the population fraction of the bright state $\ket{\uparrow}$.

The excitation with a weak optical pulse results in a mixture of four pure states $$\rho_e = \dyad{\psi_1}+\dyad{\psi_2}+\dyad{\psi_3}+\dyad{\psi_4},$$ where
\begin{align*}
\psi_1 &= (1-P_B)\ket{\downarrow\downarrow}\\
\psi_2 &= P_B\Bigl(e^{i\phi}P_e\ket{ee} + (1-P_e)\ket{\uparrow\uparrow} +  \sqrt{P_e(1-P_e)}(e^{i\phi}\ket{e\uparrow} + \ket{\uparrow e}) \Bigr)\\
\psi_3 &=\sqrt{P_B(1-P_B)}\Bigl(e^{i\phi}\sqrt{P_e}\ket{e\downarrow} + \sqrt{1-P_e}\ket{\uparrow\downarrow}\Bigr)\\
\psi_4 &= \sqrt{P_B(1-P_B)}\Bigl(\sqrt{P_e}\ket{\downarrow e} + \sqrt{1-P_e}\ket{\downarrow\uparrow}\Bigr)
\end{align*}

In the weak-excitation regime, ($P_e \to 0$), double excitation $\ket{ee}$ can be neglected. Furthermore, all zero-excitation terms can be discarded, as they will be annihilated by the photon detection superoperator $\mathcal{J}[\dyad{0}{1}]$ where $\mathcal{J}[A]B = ABA^\dagger$ . Then, the initial excited state is simplified to:
$$\rho_e = P_B^2P_e(1-P_e)(e^{i\phi}\ket{e\uparrow} + \ket{\uparrow e})(e^{i\phi}\bra{e\uparrow} + \bra{\uparrow e}) + P_B(1-P_B)P_e(\dyad{e\downarrow} + \dyad{\downarrow e})
$$

With $P_e \ll 1$, $P_e \approx P_e(1-P_e)$, and, disregarding normalization of the state:
$$\rho_e = P_B^2(e^{i\phi}\ket{e\uparrow} + \ket{\uparrow e})(e^{i\phi}\bra{e\uparrow} + \bra{\uparrow e}) + P_B(1-P_B)(\dyad{e\downarrow} + \dyad{\downarrow e})
$$

The term $P_B^2(e^{i\phi}\ket{e\uparrow} + \ket{\uparrow e})(e^{i\phi}\bra{e\uparrow} + \bra{\uparrow e})$ corresponds to the case of two emitters interfering perfectly. The term $P_B(1-P_B)(\dyad{e\downarrow} + \dyad{\downarrow e})$ corresponds to a solitary excited emitter in the cavity, with the other emitter in the dark state, in which case no interference takes place. The contribution of this term reduces the interference contrast, and one can see that the contrast is minimized for the maximally mixed state ($P_B = 0.5$).

With the initial condition $\rho(0) = \rho_e$, the system is evolved in time:
\begin{equation*}
\partial_t\rho = \mathcal{L}\rho = -i[H,\rho] + \sum_{L} \mathcal{D}[L]\rho,
\end{equation*}
where
\begin{equation*}
H = (\omega_0 + \Delta - \delta/2) \sigma^\dag_1 \sigma_1 + (\omega_0 + \Delta + \delta/2) \sigma^\dag_2 \sigma_2 + \omega_0(a_{\text{CW}}^\dag a_{\text{CW}} + a_{\text{CCW}}^\dag a_{\text{CCW}} ) + \Big[ g_{1} S_1^\dag \sigma_1 +  g_{2} S_2^\dag \sigma_2 + \text{h.c.} \Big],
\end{equation*}
$$L \in \{\sqrt{\gamma_1}\sigma_1, \sqrt{\gamma_2}\sigma_2, \sqrt{\kappa}a_{\text{CW}}, \sqrt{\kappa}a_{\text{CCW}}, \sqrt{\gamma_{d_1}}\sigma_1^\dagger\sigma_1, \sqrt{\gamma_{d_2}}\sigma_2^\dagger\sigma_2\},$$
where $\gamma_{d_i}$ is the pure dephasing rate of emitter $i$; and
$$\mathcal{D}[L]\rho = L\rho L^\dagger - \frac12(L^\dagger L\rho + \rho L^\dagger L).$$

The temporal photon wavepacket shape in the clockwise and counterclockwise direction is then given by the time-dependent expectation value of the cavity-decay number operators $\Tr[\kappa a^\dagger_{\text{CW}}a_{\text{CW}}\rho(t)]$ and $\Tr[\kappa a^\dagger_{\text{CCW}}a_{\text{CCW}}\rho(t)]$, respectively.

\section{Entanglement protocol between two emitters with $\phi = \pi/2$}\label{entanglement}

As discussed in the main text and illustrated in Fig.~4(a), for a pair of two-level systems coupled to a WGM resonator with a relative phase $\phi = \pi/2$, upon weak coherent excitation through the clockwise (counterclockwise) mode, the emitters scatter photons only in the clockwise (counterclockwise) direction; back-scattering is forbidden due to destructive interference. If, however, the emitters possess a fine structure in the ground state and spin-selective optical transitions, back-scattering is possible from a particular two-emitter Bell state, and a detection of a back-scattered photon heralds entanglement generation. 

Consider an emitter with an optical transition between states $\ket{\uparrow}$ and $\ket e$, as well as an additional spin state $\ket\downarrow$, which is not affected by the optical driving of the $\ket{\uparrow}\leftrightarrow\ket e$ transition. The first step of the entanglement protocol is spin initialization of both emitters (for instance in state $\ket{\uparrow\uparrow}$), followed by spin control to prepare each emitter in an equal superposition state
$$\ket{\psi_0} = \frac12(\ket{\uparrow} + \ket{\downarrow})_A(\ket{\uparrow} - \ket{\downarrow})_B.$$

Selective excitation of the emitters' bright spin state with a weak optical pulse produces the state
\begin{equation*}
    \ket{\psi_e} =\frac12\Bigl(e^{i\pi/2}\sqrt{P_e}\ket{e} + \sqrt{1-P_e}\ket{\uparrow} + \ket{\downarrow}\Bigr)_A\Bigl(\sqrt{P_e}\ket{e} + \sqrt{1-P_e}\ket{\uparrow} - \ket{\downarrow}\Bigr)_B,
\end{equation*}
where $P_e$ is the excitation strength, corresponding to the probability of preparing an emitter into the excited state. Expanding:
\begin{multline*}
     \ket{\psi_e} =\frac12\Bigl(e^{i\pi/2}P_e\ket{ee} + (1-P_e)\ket{\uparrow\uparrow} - \ket{\downarrow\downarrow} +  \\ \sqrt{P_e(1-P_e)}(e^{i\pi/2}\ket{e\uparrow} + \ket{\uparrow e}) +  \sqrt{P_e}(-e^{i\pi/2}\ket{e\downarrow} + \ket{\downarrow e}) + \sqrt{1-P_e}(\ket{\downarrow\uparrow}-\ket{\uparrow \downarrow}) \Bigr)
\end{multline*}
We denote the state of the waveguide as $\ket{NM}$, where $N$ and $M$ represent the number of photons emitted backwards and forwards, respectively. We now consider the final state of the emitters and waveguide after the decay:

\begin{itemize}
    \item The state $e^{i\pi/2}P_e\ket{ee}$ will emit two photons, either both right or both left, so the final state is \\ $e^{i\pi/2}P_e(\ket{\uparrow\uparrow02} + \ket{\uparrow\uparrow20})/\sqrt{2}$. 
    \item The state $\sqrt{P_e(1-P_e)}(e^{i\pi/2}\ket{e\uparrow} + \ket{\uparrow e})$ will emit forward only, so the final state is $\sqrt{2P_e(1-P_e)}\ket{\uparrow\uparrow01}$.
    \item The state $\sqrt{P_e}(e^{-i\pi/2}\ket{e\downarrow} + \ket{\downarrow e})$ will emit backward only, so the final state is $\sqrt{P_e}(\ket{\uparrow\downarrow10} + \ket{\downarrow\uparrow10})$.
\end{itemize}
The final total state is then:
\begin{multline*}
    \Psi = \frac12\Bigl(e^{i\pi/2}P_e(\ket{\uparrow\uparrow02} + \ket{\uparrow\uparrow20})/\sqrt{2} + (1-P_e)\ket{\uparrow\uparrow} - \ket{\downarrow\downarrow} +  \\ \sqrt{2P_e(1-P_e)}\ket{\uparrow\uparrow01} -  \sqrt{P_e}(\ket{\uparrow\downarrow10} + \ket{\downarrow\uparrow10}) + \sqrt{1-P_e}(\ket{\downarrow\uparrow}-\ket{\uparrow \downarrow}) \Bigr)
\end{multline*}
A photon detector that does not discriminate photon number can be modeled by a pair of measurement superoperators corresponding to ``no click'' and ``click'': $\{\mathcal{J}[\dyad{0}], \Omega_{\text{click}} \}$, where $\Omega_{\text{click}} = \sum_{n=1}^\infty \mathcal{J}[\dyad{0}{n}]$ and $\mathcal{J}[A]B = ABA^\dagger$. 
A click on a detector monitoring back-scattered photons projects the system into the (unnormalized) state:
\begin{multline*}
    \rho_l = \frac14\Biggl(\frac{P_e^2}{2}  \dyad{\uparrow\uparrow} + P_e (\ket{\uparrow\downarrow} + \ket{\downarrow\uparrow})(\bra{\uparrow\downarrow} + \bra{\downarrow\uparrow})\Biggr) = \\
    =  \frac{P_e^2}{8}  \dyad{\uparrow\uparrow} + \frac{P_e}{4} (\ket{\uparrow\downarrow} + \ket{\downarrow\uparrow})(\bra{\uparrow\downarrow} + \bra{\downarrow\uparrow}) =  \frac{P_e}{4}\Bigl(\frac{P_e}{2}  \dyad{\uparrow\uparrow} + (\ket{\uparrow\downarrow} + \ket{\downarrow\uparrow})(\bra{\uparrow\downarrow} + \bra{\downarrow\uparrow})\Bigr)
\end{multline*}
The normalized state is then:
\begin{equation*}
    \rho_l = \frac{2}{P_e + 2}\Bigl(\frac{P_e}{2}  \dyad{\uparrow\uparrow} + (\ket{\uparrow\downarrow} + \ket{\downarrow\uparrow})(\bra{\uparrow\downarrow} + \bra{\downarrow\uparrow})\Bigr) = \alpha\dyad{\uparrow\uparrow} + (1-\alpha)(\ket{\uparrow\downarrow} + \ket{\downarrow\uparrow})(\bra{\uparrow\downarrow} + \bra{\downarrow\uparrow}),
\end{equation*}
where $\alpha = \frac{P_e}{P_e + 2}$ is the infidelity of the state. In the limit of weak excitation, $\alpha \to P_e/2$. Thus, photon detection heralds entanglement whose fidelity will scale with $1-P_e/2$, and probability of detecting a photon will scale with $P_e$, a trade-off between entanglement rate and fidelity, as in reference~\citenum{humphreys2018deterministic}.

Note that, if instead the initial state $\ket{\psi_0} = \frac12(\ket{\uparrow} + \ket{\downarrow})_A(\ket{\uparrow} + \ket{\downarrow})_B$ had been prepared, one can obtain the entangled singlet Bell state $(\ket{\uparrow\downarrow} - \ket{\downarrow\uparrow})$ heralded by the detection of a forward-scattered photon.

\bibliography{Reference}